\documentclass[preprint]{aastex}

\shorttitle{HIRES Beam-matched Images}
\shortauthors{Kerton \& Martin}

\begin{document}

%TITLE AND AUTHOR INFORMATION

\title{A Fast Technique for the Creation of Large-Scale High-Resolution IRAS
(HIRES) Beam-matched Images}

\author{C. R. Kerton}
\affil{Dominion Radio Astrophysical Observatory, Herzberg Institute of
Astrophysics,\\ National Research Council, P. O. Box 248, Penticton,
BC, V2A 6K3, Canada}
\email{charles.kerton@nrc.ca}

\and

\author{P. G. Martin}
\affil{Canadian Institute for Theoretical Astrophysics and Department of
Astronomy, \\ University of Toronto, 60 St. George Street, Toronto, ON, M5S
3H8, Canada}
\email{pgmartin@cita.utoronto.ca}

%ABSTRACT

\begin{abstract}

HIRES processing provides a significant
improvement in both resolution and image quality over previous IRAS
image products, but the characteristics of the HIRES beam make accurate
comparisons between the various IRAS bandpasses and between HIRES data
and data at other wavelengths non-trivial. We present a new,
fast technique for the construction of HIRES beam-matched images that is
especially well suited for the creation of large-scale (several square
degrees) ratio maps. Other techniques for the construction of ratio
maps are discussed and compared with the new algorithm.  Examples of
the large-scale ratio maps that can be constructed using this new
technique are presented.  The algorithm's application to
the construction of multiwavelength difference images and multi-colour images
is also demonstrated.

\end{abstract}

%KEYWORDS

\keywords{techniques: image processing --- infrared: general --- dust
--- surveys}

%MAIN BODY

\section{INTRODUCTION}
\label{sec:intro}

Often one would like to create beam-matched maps of IRAS images in order to
study the properties of the emitting grains and the incident radiation
field.  For example, the ratio between the IRAS 60 and 100 $\mu$m bands
provides a measure of the equilibrium dust temperature and is often
used as a proxy for the intensity of the radiation field
\citep{hel91}. The 12/100~$\mu$m ratio and 12/25~$\mu$m ratio can be
used to trace abundance variations in different grain populations
(e.g., Boulanger et al. 1990; Ryter, Puget, \& P\'{e}rault 1987). In
order to use any such ratio image for quantitative purposes it is
important that the beam shape of the two images be well matched. 
The most commonly used
IRAS image data product for such studies is the IRAS Sky Survey
Atlas (ISSA) \citep{whe94}. With the ISSA the construction of ratio
maps does not pose a significant problem since all four IRAS bands have
been brought to a common resolution of $4' \times 5'$. Drawbacks to
using these images include the presence of striping and
the relatively low resolution.  

HIRES processing\footnote{HIRES images and ancillary files are
available as single fields ($1\degr-2\degr$ in size) through the Infrared
Processing and Analysis Center (IPAC)} (Aumann, Fowler, \& Melnyk
1990) can yield much higher spatial resolution and, with the addition
of new destriping and zodiacal subtraction algorithms \citep{cao96},
superior image quality.  However, there are a number of
characteristics of the resulting images that make the creation of an
accurate ratio map non-trivial: the achieved resolution is different
in each band for a given number of iterations, the beam shape varies
between bands primarily because of the difference in the detector
sizes, and within each band the beam shape also varies as a function
of position because of the varying IRAS scan pattern on the sky
\citep{cao97,ker00}. These image characteristics mean that a simple
ratioing of the HIRES images is not satisfactory in the majority of
cases.

Various techniques have been developed to create accurate high
resolution ratio maps with HIRES data. As
will be discussed, most existing techniques for creating HIRES ratio maps
tend to be time consuming, and in some cases also require access to
the raw IRAS data. The creation of a new fast technique
is primarily motivated by the existence of two very
large HIRES data sets that cover parts of the Galactic plane. The IRAS
Galaxy Atlas (IGA\footnote{IGA images are available at
\url{http://irsa.ipac.caltech.edu/applications/IGA/}}; Cao et al.\
1997) consists of 1$^{\rm st}$ and 20$^{\rm th}$
iteration far-infrared (60 and 100~$\mu$m) images and ancillary files
covering the entire Galactic plane between $b=\pm$4.7$\degr$. The
Mid-Infrared Galaxy Atlas (MIGA\footnote{MIGA images are available at 
\url{http://www.cita.utoronto.ca}}; Kerton \& Martin 2000) is the
equivalent mid-infrared data set (12 and 25~$\mu$m)  covering $l=74\degr$ --
$148\degr$; $b = \pm6.5\degr$.  Both of these data sets have been
incorporated as large $5.12\degr \times 5.12\degr$ mosaics into the
Canadian Galactic Plane Survey (CGPS\footnote{CGPS data are available
at the Canadian Astronomy Data Centre (CADC;
\url{http://cadcwww.hia.nrc.ca})}; Taylor 1999) an international
project that is conducting a multiwavelength (21-cm HI line, 1420 and
408 MHz continuum, CO (J$=1-0$), and infrared) survey of the Galactic
plane at a common spatial resolution of $\sim 1'$.
The existence of this large uniform data set means that
high-resolution studies of dust in the ISM can now be easily done over
very large spatial scales.  

In Section~\ref{sec:previous} we review the various techniques that
have been used to create ratio maps of HIRES images.  A new technique
developed specifically for the fast creation of large-scale HIRES
ratio maps is presented in Section~\ref{sec:algorithm} and the
resulting ratio maps are examined in Section~\ref{sec:compare}. 
Section~\ref{sec:apply} presents some sample applications of the new
technique, and conclusions are given in Section~\ref{sec:conclude}.

\section{PREVIOUS RATIO MAP CREATION TECHNIQUES}
\label{sec:previous}

\subsection{Simple Division}
\label{sec:simple}

The similarity in the size and layout of the 12 and 25 $\mu$m
detectors in the focal plane array of IRAS (IRAS Explanatory
Supplement 1988) means that the resulting
beam shapes are very similar and  two-dimensional Gaussian fits to the
beams will often agree to within the fitting errors. Beam maps
showing the HIRES beam shape at various locations in the image and
tables of two-dimensional Gaussian fits to the beam shape are
available as part of the HIRES service at IPAC and are also included
in the IGA and MIGA releases.  This similarity in beam shape means that, with
care, high resolution ratio maps can be created by simply dividing the
12 and 25 $\mu$m images.  Care is required because, although the beam
shapes are very close, the actual beam shapes are not two-dimensional
Gaussians and can vary in detail because of differences in the ratio
between the point source flux and the background level during HIRES processing
 between the two images (see Section 4.1 of Kerton \& Martin 2000 for
details). As an example, Figure~\ref{fig:s151} shows two ratio
maps of the region around the HII region Sh 2-151, chosen because of the
unusual irregular cross-shaped beam pattern caused by the significant
difference in the scan angle between the two IRAS coverages of the
region (this effect is most noticeable at high ecliptic latitudes).  The top
image was constructed by simply dividing a 12 $\mu$m image from the
MIGA  by  the corresponding 25 $\mu$m MIGA image, while the bottom
image was constructed using the cross-band simulation technique (see
Section~\ref{sec:crossband} for details) to
bring the two images to a common beam shape.  In general the structure
shown in each image is the same, but note that the point sources are more
clearly defined in the bottom image because of the matching beam
shapes. For example, the feature associated with Sh 2-151 is more clearly
shown to be a point source in the beam-matched image than in the
simple ratio image. The lower image also has a less mottled background
due to the matched beams and the slightly lower resolution.
The beneficial effect of bringing the two images to the same beam
shape is seen in the way that the cross-shaped beam pattern of the
point sources seen in Figure~\ref{fig:s151} is undistorted in the cross-band
simulator ratio image while in the simple ratio map the point sources show
some non-physical structure. Most of
the point sources (stars) in the ratio image saturate and appear as
black structures since they are brighter at 12 $\mu$m than at 25 $\mu$m.
The two noticeable exceptions are the point source associated with
Sh 2-151 at $l=108.5\degr$, $b=-2.8\degr$ and the point source at 
$l=107.6\degr$, $b=-2.24\degr$ (the planetary nebula PK 107$-$2.1).  

For some purposes a simple 12/25 $\mu$m ratio map,  such as that shown
in Figure~\ref{fig:s151}, may suffice.
The user needs to closely inspect the associated beam maps to
determine if the beam shapes are close enough for their purposes.  If
so, the similar beam shape and resolution of the mid-infrared bands
provides a quick and easy way to obtain high resolution
mid-infrared ratio maps.  Note that any simple ratio map involving the
60 and 100 $\mu$m bands will not be satisfactory because of the large
difference in beam resolution and shape, and some form of beam matching will
be desirable. This is the focus of the rest of this paper.

\subsection{Convolving to Common Lower Resolution}
\label{sec:smooth}

One simple technique that is often used is to smooth the HIRES images to a
common lower resolution. For example, \citet{xu96}, in their study of
M31, convolve their HIRES images to a common $1.7'$ circular beam
(their lowest original HIRES resolution) in order to construct ratio
images.  This technique is fast and does not require access to the
original IRAS data (just the HIRES beam tables),  but suffers from the
disadvantage that one is losing much of the resolution gained by the
HIRES processing.  This can especially be a concern for the
mid-infrared bands where one can achieve very high resolution (the
original resolution of the 12 and 25 $\mu$m images from \citet{xu96}
was $0.5' \times 0.9'$).  Over very large areas, where the beam
properties are expected to change significantly, one also has to take
care to use the correct convolving kernel to smooth the beams
correctly. If the smoothing is very large, this is not a major
concern, but again one loses some of the advantage of using HIRES in
the first place.

\subsection{Variable HIRES Iteration}
\label{sec:various}

This technique takes advantage of the fact that the resolution
achieved by HIRES processing improves as the number of iterations
increases.  By stopping the HIRES processing at different iterations
for data at different wavelengths one can obtain approximately
matching resolution in different bands (e.g., see the Wang 1994 study
of the supernova remnant IC 443).  The exact beam shape can still be
slightly different so the final images are often convolved to a
circular beam to reduce this effect \citep{wan94}.  Since this technique 
requires some degree of trial and error to match the beam resolution
in the different bands it is really only suitable for mapping small
areas and for users that have easy access to both the raw IRAS archive
and the HIRES processing software (YORIC)\footnote{The archive and
software are available at IPAC and at the Canadian Institute for Theoretical
Astrophysics (CITA)}. 

\subsection{Cross-Band Simulation}
\label{sec:crossband}

The most accurate technique for matching the various HIRES beams is
called cross-band simulation \citep{fow94}.
This technique makes use of the HIRES IRAS simulator that is part
of the YORIC software package.  Using this technique, if one wanted to create a
ratio map of a 12 and 60 $\mu$m image, the simulator would be used to
scan the 12 $\mu$m HIRES image with the 60 $\mu$m detector pattern to
create simulated data for the 12 $\mu$m sky.  These ``observations''
would then be HIRES processed as if 60 $\mu$m data to create a somewhat lower
resolution version of the 12~$\mu$m image. The same process is
followed for the 60 $\mu$m data. The final result is two images at the
same resolution (in this example, each slightly poorer than the
original 60 $\mu$m HIRES resolution).  The major advantage
of this technique is that because the resulting beams are very well
matched at every location in the image, spurious results are not
introduced into the ratio maps.  This technique is very time consuming
since the HIRES algorithm is essentially run twice for each field that
needs to be processed. Simulator images can be requested from IPAC for
single fields so a user in principle does not need direct access to
the YORIC software or the IRAS data archive for single field
images. If one is interested in creating large scale images (such as
the large mosaics that can be created with the IGA and MIGA data) then
access to the YORIC software and the IRAS data archive is essential
since large scale mosaics cannot be created by simply patching
together smaller images obtained from IPAC because of the local
destriping algorithms that are used. Both the MIGA and IGA are on a
common grid and were created using a special pre-processing technique
that allows the easy creation of high-quality large-scale mosaics
(e.g., see Cao et al. 1997 or Kerton \& Martin 2000).

\section{NEW ALGORITHM}
\label{sec:algorithm}
 
Since the typical user of the MIGA, IGA, and/or CGPS data will not have access
to either YORIC or the raw IRAS data it is important to have a robust
technique that users can apply to the data that they do have in order
to construct ratio maps.  The criteria used in the development of
this new technique are that it should make use of only those data
available to the typical user, it should be fast, and it should result in
images that are close to what the cross-band simulator technique
would produce.

As mentioned above, the resolution and position angle of the HIRES beam
varies as a function of position over an image.  As part of the
standard MIGA and IGA releases two-dimensional Gaussian fits were made
to the beam shape and recorded in accompanying files. This information is also
available for individual HIRES processing requests. The basic assumption
underlying the new ratio map technique is that the HIRES beam
is well described by the two-dimensional Gaussian at the various points in the
images. Clearly this assumption works best in those regions where the
beam shape is ``well-behaved''.  In the worst case, in regions of high
ecliptic latitude, the beam shape can even have X-shaped wings due to very
large variations in the scan angle of the IRAS satellite. At these
locations, while the assumption is clearly dubious, the resulting
ratio maps turn out to still be useful.  This is because the central
part of the X-shaped beam is still well described by the Gaussian 
approximation.

We describe the operation of the algorithm on a single standard HIRES
field that one would obtain from either the IGA or the MIGA\@. In order
to build up a large scale image, one would repeat the technique for a
number of fields then mosaic the images together.  
In a standard $1.4\degr \times 1.4\degr$ field, beam
information is reported in a $7 \times 7$ array of 
points spread evenly over the image. This sampling interval does greatly
undersample the beam variation across the image, but finer sampling was
not possible because of the way the beam maps are constructed.  The
algorithm assumes that the beam information describes the beam shape
over the $0.2\degr \times 0.2\degr$ region surrounding each beam. The
validity of this assumption depends upon how little the scan pattern
changes over the  $0.2\degr \times 0.2\degr$. Users can determine this
by either looking at the beam maps or the detector track maps that are included
in the standard MIGA and IGA releases and in a typical HIRES
request.

The algorithm runs as follows. To start, one has two images, image 1
and image 2, at IRAS band 1 and band 2 respectively.  The 
major and minor FWHM and orientation for the beams in image 2 is read
from a file the user has prepared from the HIRES beam tables.  For
each beam a two-dimensional Gaussian kernel is created and convolved with a
$0.4\degr \times 0.4\degr$ region centered on that beam that has
been clipped out of image 1. The size of the area clipped out was
chosen to provide ample room to avoid edge effects when doing the
convolution. The $0.2\degr \times 0.2\degr$ center of the
convolved region is clipped out and placed in the final convolved
image. The procedure is repeated until all 25 beams in the 1$^\circ$
center of the image have been processed and the $1\degr \times
1\degr$ convolved image has been built up. Beams centered
0.1$\degr$ from the outer edge of the image are ignored since they
correspond to a 0.2$^\circ$ border that should be clipped off the final
images before mosaicing, or making  ratio maps (this buffer zone was
included in the HIRES processing to avoid problems with artificially
low detector coverage). The entire procedure is then repeated using
image 2 and the beam data for image 1.

Essentially the new algorithm is a combination
of the smoothing technique discussed in Section~\ref{sec:smooth} and
the cross-band simulator technique described in
Section~\ref{sec:crossband}. However, instead of scanning the image
with the IRAS beam shape and detector scan pattern it uses a two-dimensional
Gaussian convolution to match the beam shapes based upon the
information provided in the beam maps and tables. Also, instead of
smoothing to a markedly lower resolution it convolves the image only
to the degree necessary to match the two images. The algorithm was
implemented using IDL because of its matrix manipulation capability,
but could easily be modified to any computer language\footnote{IDL code is
available from C.R.K. upon request}. 

While this algorithm was developed for and initially applied to IRAS HIRES data
there is nothing about it that restricts its use solely to infrared data.
The algorithm could be easily applied to, for example, an infrared and
a radio data set provided sufficient information about the beam shapes was
available (see Section~\ref{sec:apply}). Large-scale radio spectral
index maps can also be created using the algorithm. A specialized
version of the algorithm (called MOSCONV, for MOSaic CONVolve) has
been developed independently by L. Higgs at DRAO to create large-scale
($\sim 5\degr \times 5\degr$) spectral index maps using the 1420 MHz
and 408 MHz data and ancillary files contained in the CGPS.

\section{TESTS \& COMPARISONS}
\label{sec:compare}

The algorithm was tested on two regions with markedly different HIRES
beam shapes. In each case three sets of images and beam maps were created:
regular HIRES, cross-band simulated HIRES, and HIRES processed with
the new algorithm. The images allowed us to compare the relative quality
of the resulting ratio maps while the beam maps allowed us to compare in
more detail how the beam shape changes in each case. For the beam
comparisons we focused on the 12 and 100~$\mu$m HIRES beams since they
have the greatest difference in resolution and beam shape.

First we looked at a field (centered at $l = 189.0\degr$, $b =
3.0\degr$ near IC~443, see Figure~\ref{fig:ic443}) where the HIRES
beam is ``regular'', i.e. it is very close to a two-dimensional Gaussian.  

Figure~\ref{fig:regcomp} shows one-dimensional cuts through the
minor axis (at the same position angle) of the 12 and 100~$\mu$m beams
located at $l = 189.0\degr$, $b = 3.0\degr$ along with the absolute
value of the difference between the two beams.  The beam profiles have
been normalized to the same central intensity so that differences in
the beam shapes can be highlighted.  Notice the large difference in the
beam shapes in the raw images (top profile in
Figure~\ref{fig:regcomp}).  Negative rings are evident in the
100~$\mu$m beam; this is a well known artifact of HIRES processing
\citep{cao97}. The middle profiles show the beams that are created
using the algorithm presented in this paper. The match between the
beams is greatly improved from the original images. Finally the lower
cuts show the beams that are formed in the cross-band simulation
process. There is a significant improvement over the regular HIRES
beams  and even the ringing around the center of the beam is well
matched in this case.  

To illustrate the quality of the match in two dimensions, the upper
row of Figure~\ref{fig:allsur}  shows surface plots
of the absolute difference between the 12 and
100~$\mu$m beams for the regular HIRES, convolution, and cross-band 
simulation processes for this region. Again there is a substantial, comparable
improvement in the match between the beams evident in the convolution
and cross-band simulation plots. The differences
between the two beams for the cross-band simulation and convolution
beams are of the same (small) magnitude.

The same tests were repeated in another field (centered at
$l = 93.0\degr$, $b = 4.0\degr$, see Figure~\ref{fig:xbeam}) where the
HIRES beam has X-shaped wings (especially evident at 12~$\mu$m).  
The one-dimensional cuts through the various beams are shown in
Figure~\ref{fig:xcomp}. Again we see a
significant comparable improvement in the agreement between the beam
shapes both in the cross-band simulation and our new technique. The bottom
row of Figure~\ref{fig:allsur} shows the surface plots of
the difference in the beam shape. These plots
show that the convolution and cross-band simulation technique will provide
similar degrees of improvement to the quality of the beam match.

Figure~\ref{fig:allsad} shows the sum of the absolute difference
between the 12~$\mu$m and 100~$\mu$m beams for all of the beams in the
two areas we tested.  The beam that has been discussed in detail is
indicated on the figure, and clearly is a representative result.  In
all cases we see that the convolution algorithm yields a substantial
improvement in the match between the two beams that is comparable to
the results obtained using the cross-band simulation technique.

The resulting ratio maps created using our convolution technique and
the cross-band simulation technique are shown in the bottom two frames
of Figures~\ref{fig:ic443} and \ref{fig:xbeam}. The only major
difference between the two results is the detailed structure close to
the point sources. The convolution map is much faster to create than
the cross-band simulation map and results in a final ratio map that is
comparable in  achieved resolution and overall image quality.  

In order to compare the various ratio maps the rms fractional
difference between the cross-band simulated and the unprocessed HIRES
or convolution ratio image was calculated.  This was also carried
out for four subregions corresponding to various features in each
image. Results from the comparison
are presented in Tables~\ref{tbl:xb1} and \ref{tbl:xb2} for various
IRAS band combinations.  As would be expected the convolved ratio map is a much
better match to the simulated ratio map as indicated by the lowering
of the rms fractional difference for all of the various IRAS band
combinations. This quantitative assessment was combined
with a visual inspection of the convolved ratio images and the simulated
ratio images to make sure they were comparable overall.

\section{SAMPLE APPLICATIONS}
\label{sec:apply}

In Figure~\ref{fig:w5} we show an example of how this technique can be
applied to the creation of large-scale infrared ratio maps. The upper image
is the 12/100~$\mu$m ratio map of the W5 HII region created using 
ISSA images. The poor resolution and image quality (striping) referred
to in Section~\ref{sec:intro} are apparent.  The lower image is the
same image constructed using HIRES data from the IGA and MIGA
processed using our algorithm to match the 12 and 100 $\mu$m beam
shape at all locations in the image.  Clearly this image is more
useful for any scientific investigation.

The region with a low 12/100 $\mu$m ratio that dominates the center of
the map corresponds to the ionized gas in W5. The drop in this ratio
relates to the  destruction of PAH dust grains in the ionized regions
which in turn causes a  reduction in the 12~$\mu$m emission. The various dark
points visible in this map are all HII regions as confirmed by
inspection of the 1420 MHz continuum images in the CGPS database. Maps
like this one can be easily compared with large scale data products at
other wavelengths in order to investigate the properties of
interstellar dust in different environments. Note that whenever
anything unusual is found in a HIRES color map, no matter how the map
was created, a detailed examination of the separate-band images is
called for (e.g., aperture photometry, checking the noise and coverage
maps, etc.).

Although we have focused on the construction of infrared ratio maps in this
paper, another application of the new algorithm is the comparison of
multiwavelength data sets using difference images or multi-colour imaging.  In
these applications the two datasets are first scaled and then either
subtracted or displayed as different colours.  This allows the
relative distribution of the emission to be easily studied. 

Figure~\ref{fig:mw1} shows two of the typical $5.12\degr \times
5.12\degr$ mosaics from the CGPS database in the infrared (12~$\mu$m
HIRES; left panel) and in 1420 MHz continuum (right panel). The object that
dominates the center of the mosaic is LBN $140.77-1.42$. Comparison
with CO and HI images of this region in the CGPS data base show that
this region is an edge-on view of an HII-HI-molecular gas interface (a
photodissociation region or PDR).
We were interested in investigating the relative position of the
12~$\mu$m emission and the 1420~MHz continuum emission at this
interface. The beam shape in the 1420~MHz mosaic is also variable
across the mosaic, so we created beam-matched images using the beam
information provided in the CGPS database and the algorithm presented
in this paper. Lower and upper data limits (peak 95\% of image histogram)
were selected that resulted in similar visual contrast in each image
(as shown in Figure~\ref{fig:mw1}). Each image was then rescaled from
0 to 256 and the 12~$\mu$m image was subtracted
from the 1420 MHz continuum image.  The center panel of
Figure~\ref{fig:mw1} shows the resulting  difference map. Dark areas
are regions where the IR emission dominates, light areas are where
1420 MHz continuum emission dominates.  Instead of subtracting the
images one could also construct a two-colour image with the infrared
data in the red channel and radio in the blue channel (for
example). In this case we see that the intense 12~$\mu$m emission is
coming mostly from a region just to the east of the intense continuum
emission with a very slight overlap area. The PDR nature of the object
is best seen by examining a cut through the region from east to west.
Figure~\ref{fig:mw1_cut} shows a cut (background subtracted and
normalized)  from $l=141.255\degr$, $b=-0.860\degr$ to
$l=140.130\degr$ , $b=-0.965\degr$. The 12~$\mu$m emission peaks just
inside of the 1420~MHz emission and then falls off in intensity as one moves
farther into the neutral/molecular material to the east.  This rapid
increase in infrared emission at the ionized-neutral interface
followed by a slow decline in emission as one moves into the
neutral/molecular material is what one expects to see in a
cross-section of a PDR \citep{tei93}.

\section{CONCLUSIONS}
\label{sec:conclude}

Beam-matched maps of HIRES IRAS data products have been shown to be
useful both in studies of dust properties and in multiwavelength
analyses of the dust and gas phases of the interstellar medium.

For images in any of the IRAS bands, simple ratio maps  are seldom
sufficient for any application beyond superficial inspection. For
quantitative studies involving HIRES data some degree of extra
processing is required. 

To create high-resolution infrared beam-matched maps a simple, fast,
convolution technique that makes full use of the beam shape
information reported in the MIGA and IGA releases and also available
as part of the standard HIRES request output has been developed. The algorithm
provides a good match of the HIRES beams combined with a minimal loss
of the resolution gained from using the HIRES processing technique.  

We expect that this technique will be quite useful for the creation of
very large scale (over many square degrees) infrared ratio maps 
for studies of the properties of dust in the ISM.  The technique is
also useful for the construction of multiwavelength difference or
multi-colour images, and large-scale radio spectral index maps.

\acknowledgments

C.R.K. would like to thank the Ontario Graduate Scholarship program for
financial support during part of this study. The research of P.G.M. is
supported through grants from the Natural Sciences and
Engineering Research Council of Canada (NSERC).

%REFERENCES

%FIGURE CAPTIONS

\clearpage

%1

\figcaption[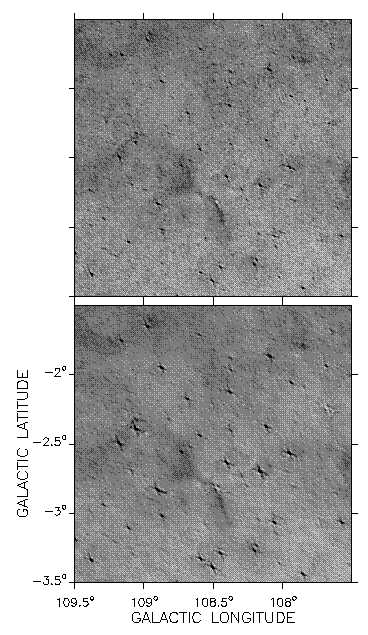]{Sh 2-151 Region. Comparison of ratio maps
(12/25 $\mu$m) constructed using a simple division of MIGA images
(top) and using the HIRES IRAS Simulator to perform cross-band beam
matching (bottom).  Images are linearly stretched from 0 (white) to 1
(black). \label{fig:s151}}

%2

\figcaption[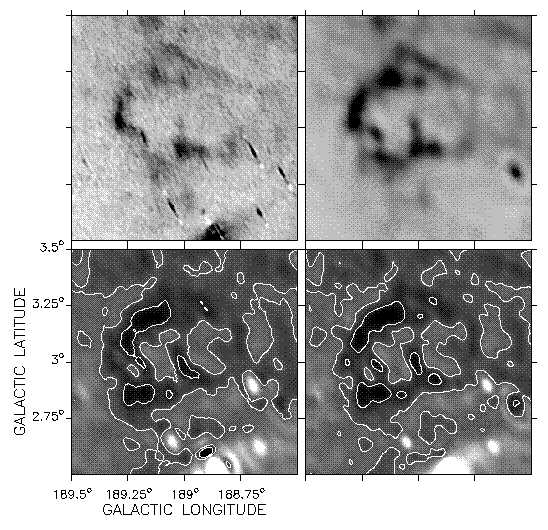]{IC 443 Region. Comparison of ratio maps 
(12/100 $\mu$m) constructed using the HIRES simulator (bottom left)
and using the convolution technique  (bottom right) for a region with
a regular HIRES beam (linear greyscale from 0.05 -- 0.14,
black--white with contours at 0.06 and 0.08). The top row shows the
original HIRES images at 12 $\mu$m (left; linear greyscale from 2.1 --
9.2 MJy/sr, white -- black) and 100 $\mu$m (right; linear greyscale
from 10 -- 120 MJy/sr, white -- black). \label{fig:ic443}}

%3

\figcaption[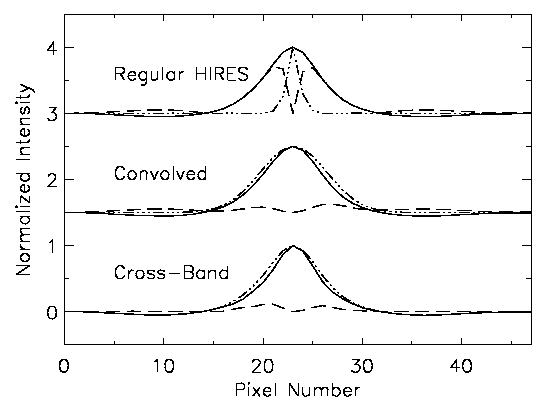]{Regularly shaped HIRES beam comparison.
The solid and dot-dash lines show normalized cuts through the minor
axis of a HIRES beam (100 and 12 $\mu$m respectively) located at
$l=189.0\degr$, $b=3.0\degr$. The dashed line shows the absolute
difference between the two beams. The regular HIRES and convolution
results have been offset by 3 units and 1.5 units respectively. 
\label{fig:regcomp}}

%4

\figcaption[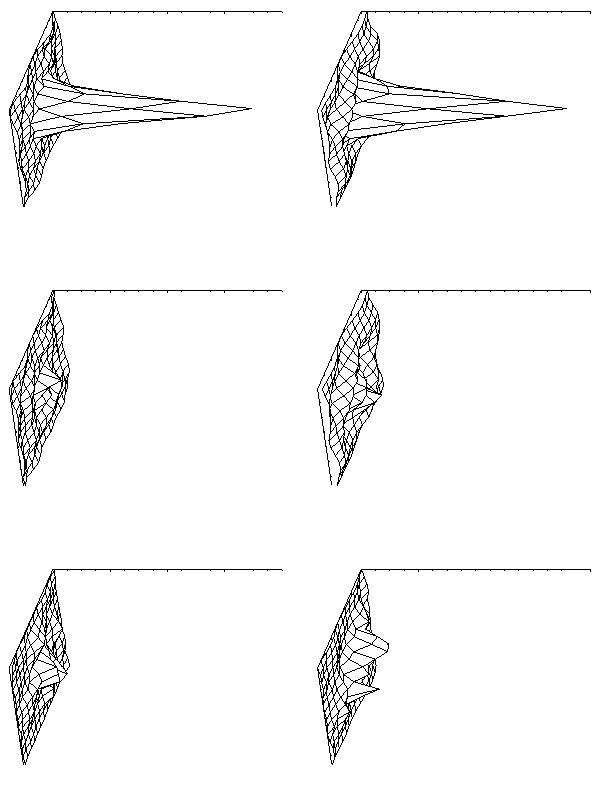]{HIRES beam absolute
differences. The top row shows the absolute difference between the
normalized 100 $\mu$m and 12 $\mu$m beams located at $l=189.0\degr$ ,
$b=3.0\degr$. The bottom row shows the results for the irregular beams
located at $l=93\degr$ ,  $b=+4.0\degr$. The differences for regular
HIRES beams, the new convolution technique and cross-band simulation
are shown from left to right.  All of the surface plots are at the
same scale. The beam images have been subsampled by a factor of 3 for
clarity.   \label{fig:allsur}}

%5

\figcaption[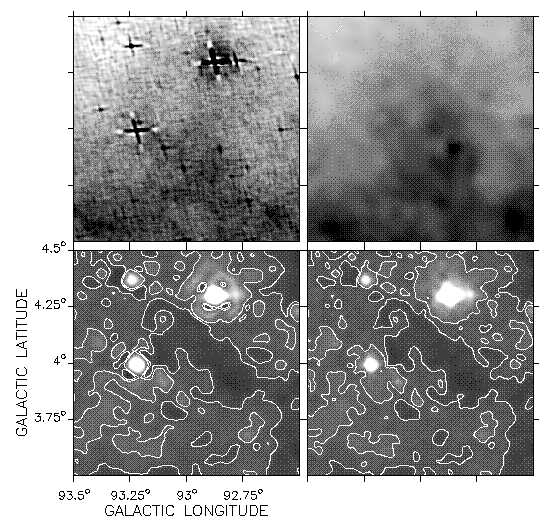]{High Ecliptic Latitude Region. Comparison
of ratio maps (12/100 $\mu$m) constructed using the HIRES simulator
(bottom left) and using the convolution technique (bottom right) for
a region with a HIRES beam with X-shaped wings (linear greyscale from
0.03 -- 0.09, black--white with contours at 0.049 and 0.053). The top
row shows the original HIRES images at 12~$\mu$m (left; linear
greyscale from 3 -- 7 MJy/sr, white -- black) and 100 $\mu$m (right;
linear greyscale from 50 -- 135 MJy/sr, white -- black). \label{fig:xbeam}}

%6

\figcaption[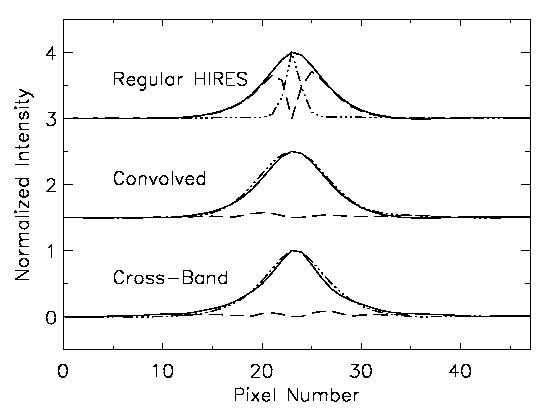]{High ecliptic latitude HIRES beam comparison.
Like Figure~\ref{fig:regcomp} but for HIRES beams located at
$l=93.0\degr$ ,  $b=+4.0\degr$. \label{fig:xcomp}}

%7
\figcaption[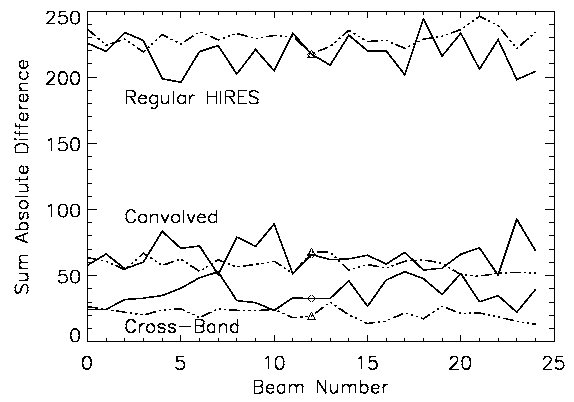]{The sum of the absolute difference between the
12 and 100~$\mu$m beams for all of beams in the two areas examined in
this paper.  The solid line is for the regularly-shaped beams and the
dot-dash line is for the high ecliptic latitude beams that have
X-shaped wings. The symbols indicate the beams that were used for the
detailed comparisons shown in Figures~\ref{fig:regcomp},
\ref{fig:allsur}, and \ref{fig:xcomp}. \label{fig:allsad}}

%8
\figcaption[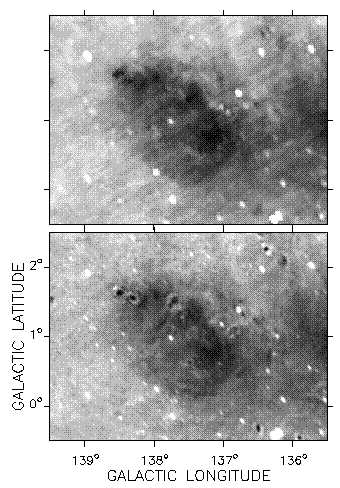]{W5 12/100 $\mu$m ratio maps.  Upper
ratio map was constructed using ISSA images, lower ratio map was
created using IGA and MIGA images processed with the algorithm
described in this paper. Both maps are linearly stretched from 0.02 --
0.08 (black -- white).  Note the significant improvement in the image
quality obtained by using the IGA and MIGA images.  Dark areas (both
extended and point source) correspond to regions of ionized gas where
PAH grains are destroyed by the hard UV radiation field.  \label{fig:w5}}

%9
\figcaption[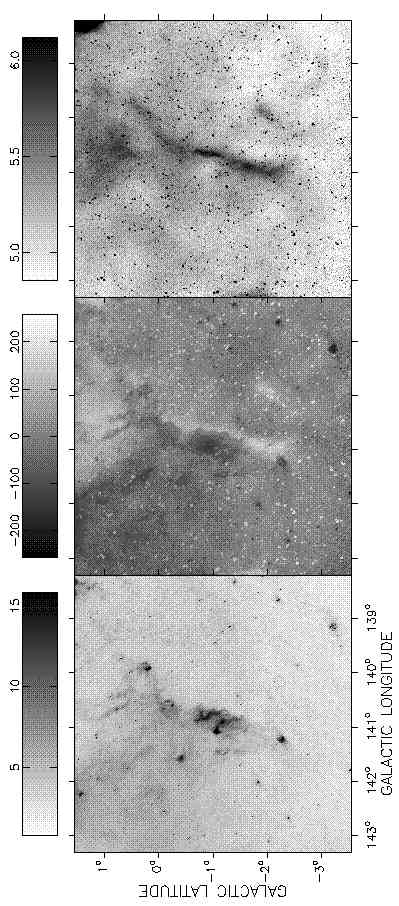]{Large-scale multiwavelength comparison.
A $5.12\degr \times 5.12\degr$ mosaic from the CGPS data base centered on
$l=140.75\degr$, $b=-1.0\degr$.  The left and right panels show the original
12~$\mu$m ([MJy/sr]) and 1420 MHz ([K]) continuum emission mosaics
respectively. The middle panel shows the result of scaling the beam-matched 
images from 0 to 256 and subtracting the infrared image from the radio image. 
\label{fig:mw1}}

%10
\figcaption[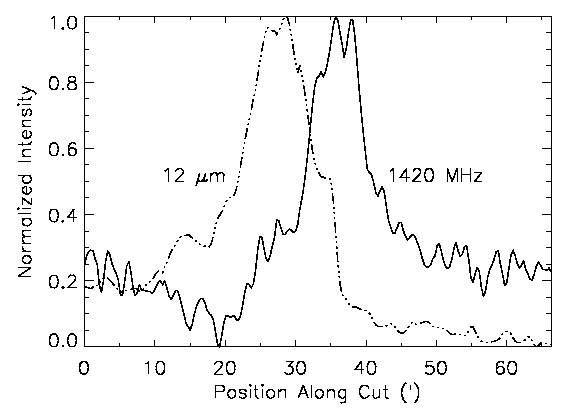]{Infrared and radio continuum emission through
the PDR LBN $140.77-1.42$ (from $l=141.255\degr$ , $b=-0.860\degr$ to
$l=140.130\degr$ , $b=-0.965\degr$ ; $18''$ wide; beam $\sim 1'$). 
The $\sim 1\degr$ long slice was taken from beam matched versions of
the images shown in Figure~\ref{fig:mw1}. Note the offset in the infrared and
radio emission peaks and the sharp increase in the infrared emission
followed by a more gradual decline. \label{fig:mw1_cut}}

%TABLES

\clearpage

\begin{deluxetable}{lccccc}
\tablecaption{Ratio Image Comparison --- Regular Beam\label{tbl:xb1}}
\tablewidth{0pt}
\tablehead{
RAW$-$SIM & &  \multicolumn{3}{c}{RMS Fractional Difference} \\
Image\tablenotemark{a} & Area 1 & Area 2 & Area 3 & Area 4 & All
}
\startdata
b1/b2 & 0.075 & 0.078 & 0.096 & 0.065 & 0.080 \\
b1/b3 & 0.084 & 0.139 & 0.162 & 0.094 & 0.100 \\
b1/b4 & 0.088 & 0.232 & 0.247 & 0.110 & 0.130 \\
b2/b3 & 0.061 & 0.074 & 0.128 & 0.105 & 0.085 \\ 
b2/b4 & 0.058 & 0.077 & 0.112 & 0.077 & 0.090 \\
b3/b4 & 0.042 & 0.047 & 0.150 & 0.090 & 0.072 \\
\cutinhead{CON$-$SIM}
b1/b2 & 0.041 & 0.050 & 0.064 & 0.040  &  0.049 \\
b1/b3 &  0.033 & 0.067 & 0.078 & 0.073  &  0.079 \\
b1/b4 & 0.013 & 0.056 & 0.061 & 0.071  &  0.098 \\
b2/b3 &  0.032 & 0.049 & 0.094 & 0.084  &  0.055 \\
b2/b4 &  0.013 & 0.030 & 0.051 & 0.057  &  0.044 \\
b3/b4 &  0.021 & 0.023 & 0.074 & 0.057  &  0.038 \\
\enddata
\tablenotetext{a}{b1 -- 12 $\mu$m, b2 -- 25 $\mu$m, b3 -- 60 $\mu$m, b4 -- 100 $\mu$m} 
\tablecomments{RAW -- no special processing; CON -- convolution
technique; SIM -- cross-band simulation \\
Area 1 -- featureless, Area 2 -- around point
source visible at 12 and 25 $\mu$m, Area 3 -- around point source
visible in all bands, Area 4 -- edge of extended structure, All ---
entire image}
\end{deluxetable}

\clearpage

\begin{deluxetable}{lccccc}
\tablecaption{Ratio Image Comparison --- Irregular Beam\label{tbl:xb2}}
\tablewidth{0pt}
\tablehead{
RAW$-$SIM & &  \multicolumn{3}{c}{RMS Fractional Difference} \\
Image\tablenotemark{a} & Area 1 & Area 2 & Area 3 & Area 4 & All
}
\startdata
b1/b2 & 0.030 & 0.109  & 0.191 & 0.030 & 0.060 \\
b1/b3 & 0.038 & 0.161  & 0.280 & 0.043 & 0.087 \\
b1/b4 & 0.046 & 0.305  & 0.431 & 0.058 & 0.131 \\
b2/b3 & 0.020 & 0.093  & 0.247 & 0.030 & 0.054 \\ 
b2/b4 & 0.022 & 0.165  & 0.396 & 0.034 & 0.085 \\
b3/b4 & 0.017 & 0.025  & 0.138 & 0.065 & 0.032 \\
\cutinhead{CON$-$SIM}
b1/b2 & 0.016  & 0.089 & 0.139 & 0.015 & 0.041 \\
b1/b3 & 0.012  & 0.085 & 0.272 & 0.021 & 0.055 \\
b1/b4 & 0.010  & 0.074 & 0.241 & 0.016 & 0.049 \\
b2/b3 & 0.010  & 0.035 & 0.192 & 0.022 & 0.037 \\
b2/b4 & 0.007  & 0.032 & 0.160 & 0.015 & 0.031 \\
b3/b4 & 0.007  & 0.008 & 0.044 & 0.017 & 0.011 \\
\enddata
\tablenotetext{a}{b1 -- 12 $\mu$m, b2 -- 25 $\mu$m, b3 -- 60 $\mu$m, b4 -- 100 $\mu$m} 
\tablecomments{RAW -- no special processing; CON -- convolution
technique; SIM -- cross-band simulation \\
Area 1 -- featureless, Area 2 -- around point
source visible at 12, 25, and 60 $\mu$m, Area 3 -- around very bright point source visible at 12, 25, and 60 $\mu$m, Area 4 -- around point source visible in all bands, All --- entire image}
\end{deluxetable}

\end{document}